# A Multi-frequency Magnetic Particle Spectroscopy System for the Characterization of Magnetic Nanoparticles

Shaoqi Sun, Shijie Sun, Lijun Xu, and Jing Zhong

*Abstract*—Magnetic particle spectroscopy (MPS) is one of the most versatile methods to characterize the magnetic properties of magnetic nanoparticles (MNPs). The excitation magnetic field is one of the most crucial factors that affects the MPS signal of the MNPs. In this study, a multi-frequency MPS system is developed to investigate the MPS signal of MNPs in different ac magnetic fields. The MPS system consists of a multi-channel excitation module for the generation of different-frequency ac magnetic fields and a detection module for the measurement of the magnetic response of the MNPs. The MPS system allows to generate ac magnetic fields with a frequency up to 32.6 kHz and amplitude up to 25 mT. The MPS signals of the MNPs in different ac magnetic fields are measured to systematically evaluate the performance of the multi-frequency MPS system, including the MNP spectra and its dynamic magnetization curve. In addition, the signal-to-noise ratio (SNR) of the MPS system is quantitively assessed with measured MPS signals of a given MNP sample and DI water. Furthermore, a series of MNP samples with different iron concentrations are prepared and measured to evaluate the limit-of-detection (LOD) in terms of iron concentration. The influence of the excitation magnetic field, including frequency and amplitude, is discussed based on the SNRs of the measured harmonics. Experimental results show that the LOD is 2.3 ng in terms of iron.

## I. INTRODUCTION

Magnetic nanoparticles (MNPs) have been widely used as emerging platforms in biomedical applications, including heaters in magnetic hyperthermia [1]–[3], carriers in targeted drug delivery [4]–[6], tracers in magnetic particle imaging (MPI) [7]–[13] and sensors in magnetic bio-sensing [14]–[17]. In magnetic hyperthermia, MNPs are remotely heated when exposed to a radio-frequency magnetic field, allowing to kill tumor cells for cancer therapy. In targeted drug delivery, MNPs are encapsulated together with drugs into polymers for the delivery of the drugs to a specific tumor tissue. In addition, MNPs have been introduced as tracers in MPI [9], which allows for fast, and quantitative MNP imaging. In magnetic bio-sensing, functionalized MNPs are used as sensors for the detection of biomarkers, e, g. antigen, antibody, and virus [14], [16], [18]–[20]. In these biomedical applications, the magnetic properties of the MNPs play significant roles. Thus, it is of great importance to characterize the MNP magnetic properties, including the magnetization spectra and dynamic magnetization curve.

Magnetic particle spectroscopy (MPS) system is a versatile tool for the characterization of MNPs' magnetic properties. To date, there have been several approaches reporting on the design of different-configured MPS systems for the characterization of MNPs, as well as for biomedical applications. For instance, Garraud *et al*. designed an MPS system to measure the linear dynamic magnetic susceptibility at 1 mT with a frequency range from 0.5 Hz to 120 kHz and the nonlinear spectra at 50 mT with resonant and matching circuits at discrete frequencies from 3 kHz to 24 kHz [21]. Behrends et al. developed a frequency-tunable MPS system containing 4 measurement ranges base on impedance matching [22], which showed the ability to perform MPS measurements in the range of 100 Hz-24 kHz. Draack et.al developed a multiparametric MPS system applied on $CoFe_2O_4$ nanoparticles in viscous media and realized the multi-parameter measurement characterizing dynamic magnetic behavior under 25 mT [23]. Löwa et al. developed a benchtop MPS system applied for monitoring the dynamic magnetization behavior of MNPs during the synthesis at a fixed excitation frequency of 25 kHz and reached the detection limit of 1.4 ng [24] .Wu et al. designed a portable handled MPS device based on mixing-frequency detection method, which reached the limit of detection (LOD) of 4 μg of iron oxide MNPs. Afterwards, the MPS device was applied to wash-free magnetic bioassays for the detection of streptavidin[15], [25]–[27]. Zhong et al. investigated MNP-based biomolecule (MNPs conjugated by the biotinylated Immunoglobulin G, IgG) detection with an MPS system on the MNP concentration, quantitatively studied the relaxation time, MPS spectra of the IgG conjugated MNPs [20]. Graeser et al. designed a 2-dimensional MPS system with 2D excitation magnetic fields and 2D detection coils for the detection of MNPs, reaching a limit of detection of 450 pg Fe [28]. In an MPS system, the generation of different-frequency magnetic fields with sufficiently large amplitude is of great importance to the characterization of the MNPs. Additionally, the limit of detection, determined by signal-to-noise ratio, plays a key role in the measurement accuracy of the magnetic response of the MNPs.

In this paper, we developed a multi-frequency MPS system to characterize the magnetic properties of MNPs in different

This work was supported by National Natural Science Foundation of China (62271025 and 62027901), the Key-Area Research and Development Program of Guangdong Province (2021B0101420005) and the Fundamental Research Funds for the Central Universities.

Shaoqi Sun, Lijun Xu, Shijie Sun and Jing Zhong are with the School of Instrumentation and Opto-Electronic Engineering, and Key Laboratory of Precision Opto-mechatronics Technology of Education Ministry, Beihang University, Beijing 100191, China. (*Corresponding author, Jing Zhong, zhongjing@buaa.edu.cn).

ac magnetic fields. At first, the MPS principle and the multi-frequency MPS system design are introduced. Afterwards, the dynamic magnetization curve and its spectra of the MNPs are measured while the signal-to-noise ratio (SNR) of the MPS system is quantitatively and comprehensively investigated. A series of diluted MNP samples with different iron concentration is used to determine the LOD in terms of iron mass. The influence factors that affect the LOD are discussed.

## II. SYSTEM DESIGN

### A. MPS principle

MPS directly measures the magnetic response of MNPs induced in an ac magnetic field. The frequency of an excitation magnetic field is one of the key factors affecting the dynamic magnetization and its spectra of the MNPs. It is of great importance to investigate the magnetic response of the MNPs induced in ac magnetic fields with different frequencies and amplitudes. Fig. 1 shows the schematic of the MPS principle. As shown in Fig. 1a, ideally, the static magnetization curve of the MNPs can be described by the static Langevin function. When exposed to a time-varying magnetic field, the magnetic response of the MNPs – MNP magnetization $M(t)$ cannot fully follow the ac magnetic field, thus showing a phase lag (see Fig. 1b). It depends on the MNP relaxation, the excitation frequency and amplitude, as well as other environmental parameters. With a detection coil-based measurement system, the output voltage $V(t)$ is proportional to the time derivative of MNP magnetization $M(t)$. Due to the superparamagnetism of the MNPs, the nonlinear MNP magnetization $M(t)$ and the output voltage $V(t)$ contain not only the fundamental harmonic, but also high harmonics (see Fig. 1c), depending on the strength of the excitation magnetic field. Measuring the spectra of the MNPs allows for the reconstruction of the dynamic magnetization of the MNPs. Therefore, the measurement of the MPS signal of the MNPs, in principle, allows one to characterize the magnetic properties of the MNPs, the relaxation mechanism, and the determination of environmental parameters.

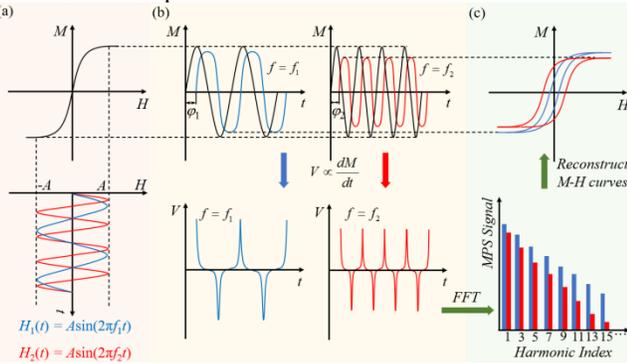

Fig. 1. The schematic of the MPS principle in different-frequency ac magnetic fields. (a) The static magnetization curve of the MNPs and the time-varying excitation of different-ac magnetic fields. (b) The time-varying magnetic moment response of MNPs and the corresponding output MPS signal of a detection-coil based measurement system. (c) Fourier transform (FFT) of the MPS signal and the reconstructed dynamic magnetization curve of the MNPs.

### B. MPS system design

Figure 2 shows the schematic of the custom-built multi-frequency MPS system, consisting of a multi-channel excitation module and a gradiometric coil-based detection module. In the multi-channel excitation module, the excitation coil is driven by a power amplifier to generate a desired ac magnetic field. Several serial capacitors are used to match the excitation coil for working at different resonance frequencies. The series capacitors are switched by a program-controlled relay. A current sensor is used to measure the current passing through the excitation coil for the measurement of the magnetic field strength. In the detection module, a gradiometric detection coil is used to measure the magnetic response of the MNPs induced in the ac magnetic field. The output signal of the gradiometric detection coil is pre-amplified and digitized for further processing in PC.

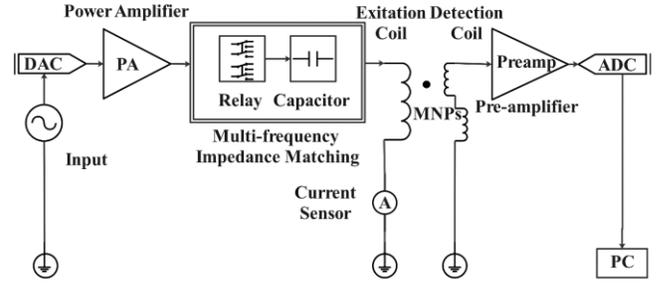

Fig. 2. Schematic of the multi-frequency MPS system.

Figures 3a and 3b show the CAD model and the photo of the excitation and detection coil systems, respectively. The coil bobbins are 3D-printed by white photosensitive resin. The excitation coil is wound with Litz wires (outer diameter $D$ = 2.21 mm, 0.1 mm × 250 strands) with 6 layers and 120 turns approximately. The program-based relay allows the switching of the capacitors at seven resonance frequencies from 5 kHz to 32.6 kHz. For excitation frequency not higher than 2.5 kHz, the power amplifier is able to drive the excitation coil without a matching capacitor.

The detection module consists of a gradiometric detection coil (receive and compensation coils) and a preamplifier. Both of the coils are wound with copper wires (diameter $D$ = 0.15 mm, 4 layers and 132 turns approximately). The receive coil measures the MNP response and the direct feedthrough from the excitation magnetic field whereas the compensation coil only measures the direct feedthrough from the excitation magnetic field. Ideally, the gradiometric detection coil only measures the magnetic response of the MNPs and suppresses the direct feedthrough from the excitation magnetic field. However, due to positioning and impedance difference of the receive and compensation coils, the output signal of the gradiometric detection coil still contains some residual feedthrough, which should be digitally subtracted.

The whole measurement procedure includes a blank measurement and an MNP measurement. A blank measurement without an MNP sample is performed at first to measure the residual feedthrough. Afterwards, an MNP sample is placed into the sample holder of the measurement system. The measured signals without and with an MNP sample are subtracted to obtain the pure signals generated by the MNPs. During the whole measurement procedure, the strength of the excitation magnetic field is controlled at a given value. The complex MPS signals are calculated by a digital lock-in amplifier, implemented in PC. Each measurement of the MPS signal takes 2 sec plus the time for placing the MNP sample.

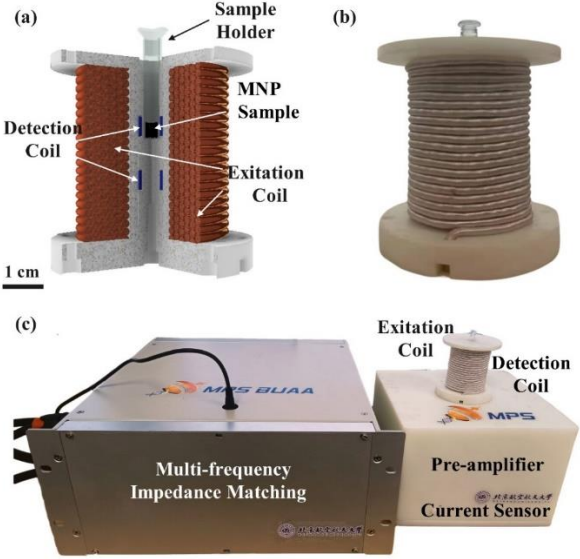

Fig. 3. Photos of the MPS system. (a) CAD model of coil unit. (b) Photography of coil unit. (c) Photography of the multi-frequency MPS system including impedance matching unit.

### III. RESULTS AND DISCUSSION

#### A. Experimental description

In this paper, the experimental sample is Synomag® D-50 Plain, purchased from micromod Partikeltechnologie GmbH (Rostock, Germany). The multi-core based MNPs have a nominal hydrodynamic size of 50 nm. The iron concentration of the stock sample is 10 mg/mL. For experiments, the volume of each sample amounts to 60 μL. The MPS system allows to generate an ac magnetic field with arbitrary frequencies up to 2.5 kHz whereas a series of capacitors are used to generate high-frequency ac magnetic field at 5 kHz, 7.36 kHz, 9.8 kHz, 14.9 kHz, 20.2 kHz, 25.35 kHz and 32.6 kHz. At all the given frequencies, the maximum strength of the excitation magnetic field is 25 mT. To determine the LOD, the noise level and SNRs are measured with an MNP sample and DI water sample in ac magnetic field with different excitation frequencies and amplitudes.

#### B. MNP Characterization

The magnetic properties of the MNPs, including the complex MPS signal and dynamic magnetization curve, are characterized in ac magnetic fields with different frequencies and amplitudes with the custom-built multi-frequency MPS system. Fig. 4 and Fig. 5 show the measured spectra and dynamic magnetization curve of the MNPs in different-frequency magnetic fields with amplitudes ranging from 10 mT to 25 mT, respectively. Fig. 4 indicates that with increasing the harmonic index, the harmonic amplitude gradually decreases. In addition, with increasing the excitation frequency $f_0$, the MPS signal decreases with increasing the harmonic index (the decay curve) faster. Comparing Figs. 4a – 4d, with increasing the amplitude of the excitation magnetic field $H_0$, the decay curve decreases more slowly. Fig. 5 shows that with increasing $f_0$, the hysteresis loss area gets larger. Due to the MNP relaxation, at a higher excitation frequency $f_0$, the MNPs cannot follow the excitation magnetic field, resulting in the faster decay curve and the larger hysteresis loss of the dynamic magnetization curve. The experimental results demonstrate that the built multi-frequency MPS system allows to measure the dynamic magnetization of the MNPs and its spectra in different ac magnetic fields.

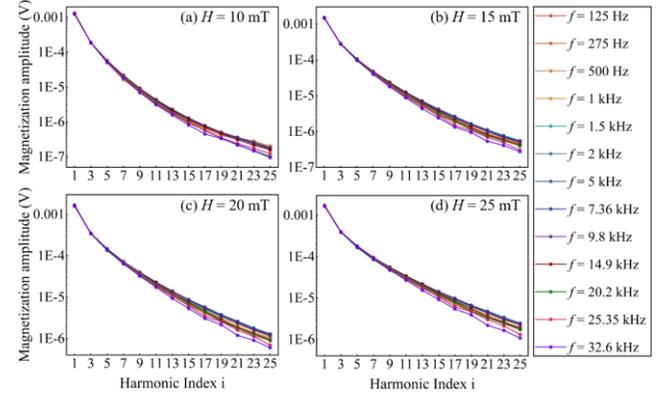

Fig. 4. Experimental results of the MNP spectra in ac magnetic fields of 10 mT (a), 15 mT (b), 20 mT (c), 25 mT (d), respectively. Symbols represent the measured data whereas solid lines are guides to eyes.

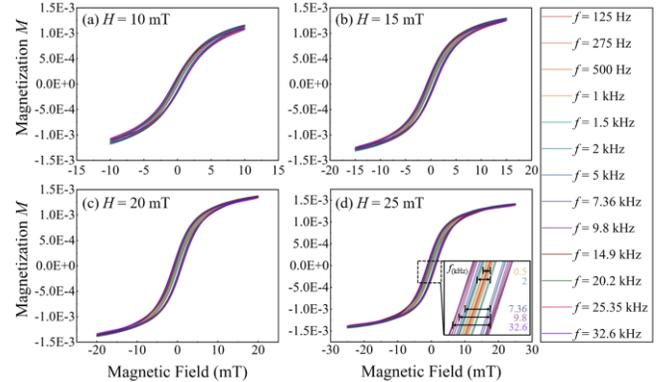

Fig. 5. Dynamic magnetization curves $M(H)$ of multi-frequency MPS. (a) – (d) figures show the measured curves excited by ac magnetic fields of 10 mT, 15 mT, 20 mT, 25 mT, respectively.

#### C. Signal-to-noise Ratio

SNR is one of the most important metrics that affects the performance of the MPS system, e.g. LOD. The MPS signal of the MNPs and the noise are quantitatively measured to evaluate the SNR of the MPS system. The noise source of an MPS system contains not just the general noise from electronics, e.g. Gaussian noise, but also the coupled interference from the harmonic distortion of the excitation magnetic fields. A DI water sample is used to measure the noise. Each measurement is repeated 10 times. Note that with an excitation magnetic field of 0.02 mT, the harmonic distortion of the excitation magnetic field can be neglected.

Thus, the measured signal of the DI water sample can be considered as the general noise from electronics. With an excitation magnetic field of high amplitudes, e.g. 10 mT, 15 mT, 20 mT and 25 mT, the harmonic distortion should be taken into account. Consequently, the measured signal of the DI water sample at the high-amplitude magnetic field are the mixture of the general noise from electronics and the coupled interference from the harmonic distortion.

Figure 6 shows the measured $i^{th}$ harmonic $M_i$ vs. excitation frequency $f_0$ ($M_i$-$f_0$) curves in ac magnetic fields with different amplitudes $H_0$ with a DI water sample. With $H_0 = 0.02$ mT, the measured $M_1$ roughly deceases with increasing the excitation frequency, which may be caused by the digital lock-in method at high frequency allowing for a better SNR. In Fig. 6a, with $H_0 = 10$ mT, 15 mT, 20 mT and 25 mT, the measured $M_1$ does not significantly changes with increasing the excitation frequency, but is higher than that with $H_0 = 0.02$ mT. It might be caused by the direct feedthrough from the excitation magnetic field, which leads to instability in $M_1$ during the blank and sample measurements. Figs. 6b-6d show the measured $M_i$ gradually decreases with increasing the excitation frequency $f_0$. It might be caused by the 1/f noise and the calculation performance of the digital lock-in amplifier at high frequencies. At some excitation frequencies, e.g. 1 kHz, 1.5 kHz and 2 kHz, the measured $M_3$, $M_5$, and $M_7$, at $H_0 = 0.02$ mT are slightly lower than those at $H_0 = 10$ mT, 15 mT, 20 mT and 25 mT. It might be caused by the harmonic distortion in the excitation magnetic field, which directly feeds to the measurement system. With excitation frequency $f_0 = 1$ kHz, 1.5 kHz, and 2 kHz, the power amplifier can directly drive the excitation coil without using series capacitors for impedance matching. However, the required powers at those frequencies are significantly higher, thus causing some harmonic distortion. In addition, the resonance circuit with impedance matching can also filter other frequency signal except at the matching frequency. Those interference from the harmonic distortion of the excitation magnetic field may worsen the LOD of the MPS system. Compared to the measured $M_1$, the measured $M_3$, $M_5$, and $M_7$ monotonically decrease as excitation frequency increases, showing less influence by feed-through coupling. Thus, noise level characterized by these harmonics is lower at higher excitation frequencies e.g. 20.2 kHz, 25.35 kHz and 32.6 kHz.

Figure 7 shows the measured SNRs of the $i^{th}$ harmonic vs. excitation frequency $f_0$ curves in ac magnetic fields $H_0$ with different amplitudes. An MNP sample with iron concentration of 1 mg/mL is used to measure the signal whereas a DI water sample is used to measure the noise, presented in Fig. 6. Fig. 7a shows that the SNR of the 1st harmonic does not significantly change with an increase in the excitation frequency. It may be caused by the instability of direct feedthrough from the excitation magnetic field during blank and MNP measurements. In contrast, the higher harmonics, e.g. the 3rd, 5th, 7th harmonics, increase with increasing the excitation frequency.

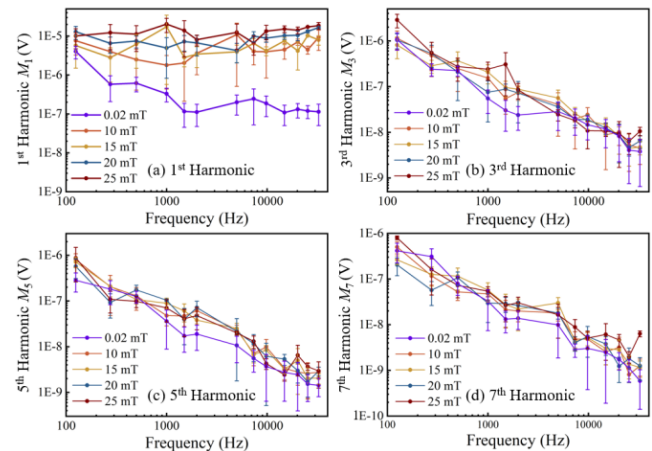

Fig. 6. The measured $i^{th}$ harmonic vs. excitation frequency curves in ac magnetic fields with different amplitudes. (a) – (d) figures show the measured 1st, 3rd, 5th, and 7th harmonics, respectively. Symbols represent the measured data whereas solid lines are guides to eyes. Error bars represent standard deviation of 10 averaged measurements.

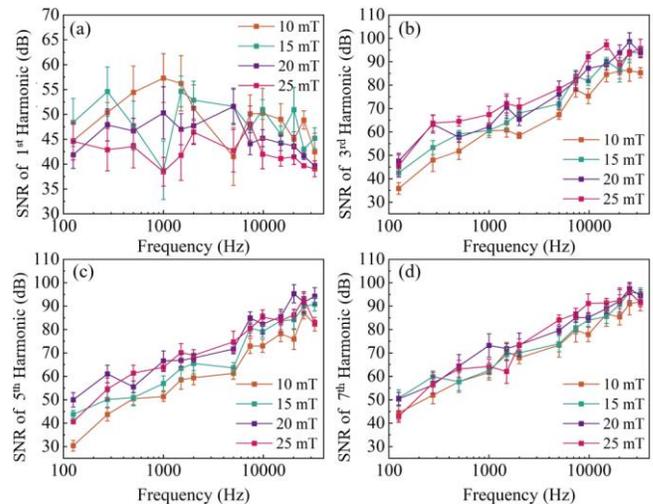

Fig. 7. The measured SNR at $i^{th}$ harmonic vs. excitation frequency in ac magnetic fields with different amplitudes while testing Synomag® D-50 Plain solution ($c$(Fe)=1mg/mL, iron mass $m$(Fe) = 60 μg). (a) – (d) figures show the SNR characterized by 1st, 3rd, 5th, and 7th harmonics, respectively. Symbols represent the measured data whereas solid lines are guides to eyes.

### D. Limit of detection

The limit of detection (LOD) is another metrics to evaluate the performance of an MPS system, which is of great importance to magnetic biosensing. To determine the LOD of our custom-built MPS system, a series of diluted MNP samples is prepared for experiments. The iron concentration of the experimental samples ranges from 10 μg/mL to 9.8 ng/mL, as well as the sample of DI water without iron. The photo of the experimental sample is depicted in Fig. 8. Note that each sample was measured 4 times.

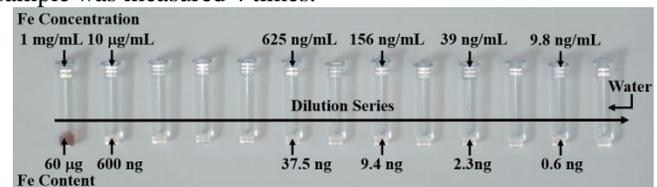

Fig. 8. Dilution series of Synomag® D-50 Plain samples used in the experiments.

The MPS signals of the MNP samples with different iron concentrations are measured in ac magnetic fields with 4 different frequencies and 4 different amplitudes. Fig. 9 shows the measured MPS signals in ac magnetic fields with frequency of 25.35 kHz whereas the MPS signals at other frequencies of 14.9 kHz, 20.2 kHz, and 32.6 kHz are presented in Figs. S1-S3 in Supporting Information (SI). With decreasing the iron concentration, the strength of the MPS signal gradually decreases, showing linear behaviour with the iron concentration. Thus, the MPS signal of the MNPs can be used to quantify the iron content. In principle, the measured MPS signal should decrease with increasing the harmonic index. However, the measured MPS signal fluctuates with harmonic index of about 23. The frequency of the harmonic is 23×25.35 kHz = 583.05 kHz. The fuctuation mainly comes from the noise at frequency range of about 583.05 kHz.

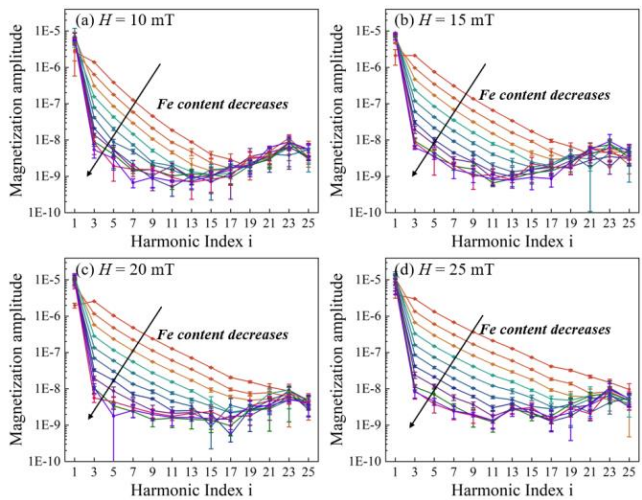

Fig. 9. Experimental results of the MPS signal of the diluted MNP samples in ac magnetic fields with frequency $f = 25.35$ kHz and amplitudes $H$ of (a) 10 mT, (b) 15 mT, (c) 20 mT, (d) 25 mT. Symbols represent the measured data whereas solid lines are guides to eyes.

Based on the analysis on the SNRs of the measured harmonics, the measured 3rd, 5th and 7th harmonics are used to calculate the measurement sensitivity $S$ of the harmonic amplitude to iron content and the LOD in terms of iron. Figs. S4-S7 in SI present the 3rd, 5th and 7th harmonic vs. iron content curves in ac magnetic fields with different frequencies and amplitudes. The measurement sensitivities $S$ is calculated by linear fitting, which is the slope of the harmonic amplitude vs. iron content curves. Figs. 10a-10c shows the calculated measurement sensitivities of the 3rd, 5th and 7th at different frequencies and amplitudes. The standard deviation $\sigma$ of the background noise, measured with the sample of the DI water, is calculated and presented in Fig. 10d-10f. They indicate that with increasing the magnetic field strength the measurement sensitivity increases. In addition, the measurement sensitivity of the 3rd harmonic is higher than that of the 5th and 7th harmonics. The standard deviation fluctuates, which depends on the background noise and the harmonic distortion as discussed above.

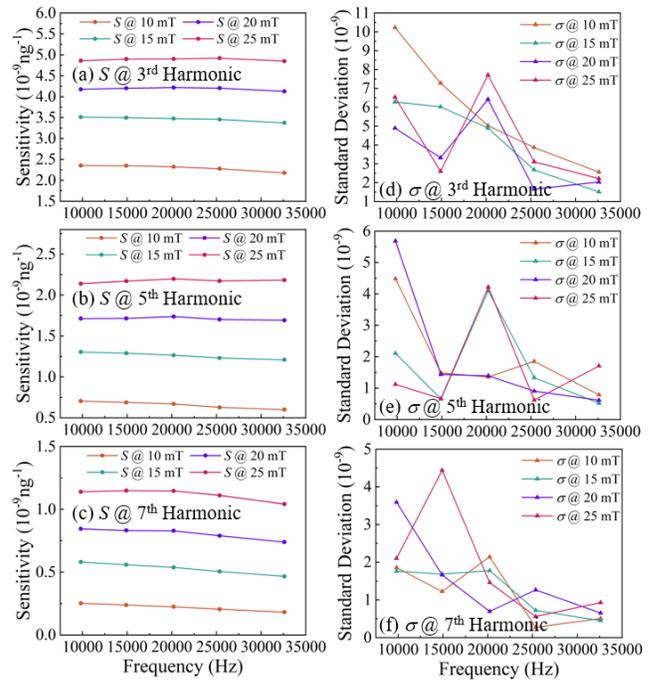

Fig. 10. The measurement sensitivity $S$ of the (a) 3rd, (b) 5th, (c) 7th harmonics to iron content and the standard deviation $\sigma$ of the noise level on the (d) 3rd, (e) 5th, (f) 7th harmonics at different frequencies and amplitudes.

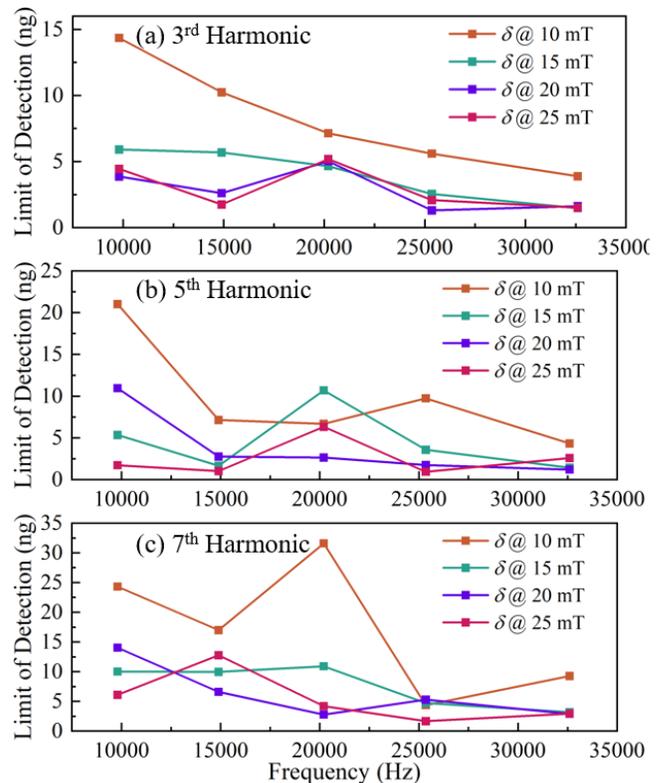

Fig. 11. The estimated LOD of the 3rd, 5th, and 7th harmonics in terms of iron.

With the calculated measurement sensitivity and the standard deviation of the background noise, one can calculate the LOD in terms of iron. Fig. 11 shows the calculated LOD $\delta$ of different harmonics in terms of iron content, estimated by $\delta = 3.3\sigma/S$. In general, the calculated LOD gets improved with increasing the excitation frequency. It's worth noting that the estimated LOD level is higher with higher harmonics (e.g. 7th harmonic) and under lower excitation field amplitudes (e.g. 10 mT), which is consistent with the analysis on SNR above. The LOD with the 3rd harmonic under 20 mT, 25 mT is in the range of approximately 1.5 ng to 5.2 ng. The best LOD is achieved at excitation field frequency of 25.3 kHz, 32.6 kHz.

Figure 12 shows the best results of the LOD with the harmonic amplitude vs. iron content at lower iron contents, which is the local zoom-in figures of the Figs. S4-S7. For instance, at frequency of 25.35 kHz, the 4-averaged harmonic amplitude of the 3rd harmonic at 25 mT represents an LOD of 2.3 ng iron. At 32.6 kHz, the 5th harmonic allows to achieve the LOD of 2.3 ng iron. Therefore, they indicate that the multi-frequency MPS system achieve an LOD of 2.3 ng in terms of iron. The LOD mainly depends on SNR of the measured harmonics of the MNP dynamic magnetization. For a given MNP sample, one can improve the LOD by decreasing the background noise of the measurement system, including Gaussian noise from electronics and the harmonic distortion.

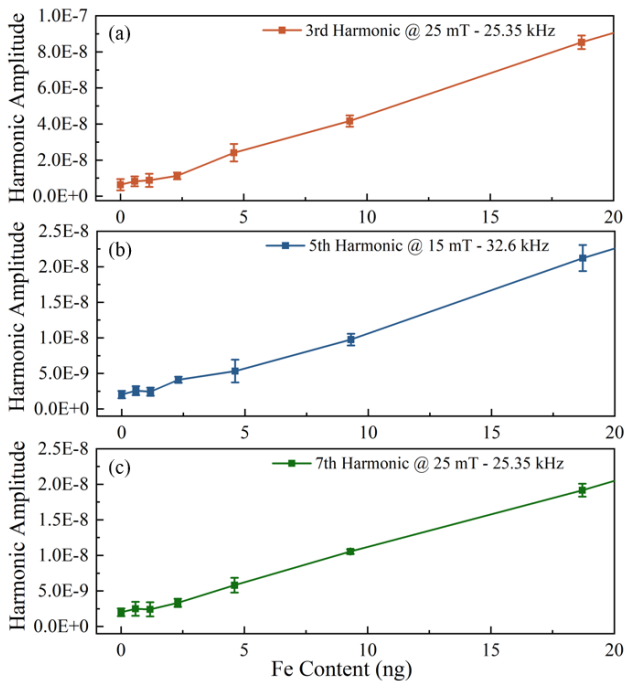

Fig. 12. (a) The 3rd harmonic $M_3$ vs. the iron content at amplitude $H$ = 25 mT and frequency $f$ = 25.35 kHz. (b) The 5th harmonic $M_5$ of the diluted MNP samples under magnetic field intensity of $H$ = 15 mT at frequency of $f$ = 32.6 kHz. (c) The 7th harmonics $M_7$ of the dilution series samples under magnetic field intensity of $H$ = 25 mT at frequency of $f$ = 25.35 kHz.

## IV. Conclusion

In this paper, we developed a multi-frequency magnetic particle spectroscopy (MPS) system for the characterization of MNPs in different ac magnetic fields. The multi-frequency MPS system consists of a multi-channel excitation module with a series of capacitors for the generation of the different-frequency ac magnetic fields and a gradiometric coil-based detection module for the measurement of the magnetic response of the MNPs. The dynamic magnetization curve and the MPS signal of the MNPs were measured with the developed multi-frequency MPS system. The signal-to-noise ratio (SNR) of the measured harmonics is quantitatively assessed with the measured MPS signals of the MNPs and DI water. In addition, the LOD of the MPS system in terms of iron with different harmonics is evaluated with a series of diluted MNP samples with different iron concentrations while the influence factors on the LOD are discussed. We envisage that the developed multi-frequency MPS system is of great importance to characterize the MNPs for magnetic particle imaging and magnetic biosensing.